\documentclass[11pt,fleqn]{article}

\usepackage{amsmath}
\usepackage{amsfonts}
\usepackage[dvips]{graphicx}
\usepackage{dcolumn}
\usepackage{upgreek}
\usepackage{latexsym,slashed}
\usepackage{amssymb,amsfonts}
\usepackage{cancel}
\usepackage{bm}        %% bold mathematics
\usepackage{amsmath,mathrsfs}   %% non gradito a latex2html!

\newcommand{\pacs}[1]{\smallskip\noindent{\sl PACS numbers:
                       \hspace{0.3cm}#1}\par\bigskip\rm}
\newcommand{\ack}[1]{\par\section*{Acknowledgments} #1}

\newcommand{\address}[1]{\begin{center}\large #1\end{center}}

\def\beq{\begin{eqnarray}}
\def\eeq{\end{eqnarray}}

\def\p{\partial}

\def\R{{\hbox{{\rm I}\kern-.2em\hbox{\rm R}}}}   %% real numbers
\def\H{{\hbox{{\rm I}\kern-.2em\hbox{\rm H}}}}   %% Hilbert space
\def\N{{\hbox{{\rm I}\kern-.2em\hbox{\rm N}}}}   %% natural numbers
\def\C{{\ \hbox{{\rm I}\kern-.6em\hbox{\bf C}}}} %% complex numbers
\def\Z{{\hbox{{\rm Z}\kern-.4em\hbox{\rm Z}}}}   %% integers numbers
\catcode`\@=11
\@addtoreset{equation}{section}
\begin{document}
\tolerance=5000

\title{Applications of the Tunneling Method to Particle Decay
  and Radiation from Naked Singularities}   

\author{Roberto~Di~Criscienzo$\,^{(a)}$\footnote{rdicris@science.unitn.it},
  Luciano~Vanzo$\,^{(a)}$\footnote{vanzo@science.unitn.it} and
  Sergio~Zerbini$\,^{(a)}$\footnote{zerbini@science.unitn.it}}
\date{}
\maketitle
\address{$^{(a)}$ Dipartimento di Fisica, Universit\`a di Trento \\
and Istituto Nazionale di Fisica Nucleare - Gruppo Collegato di Trento\\
Via Sommarive 14, 38123 Povo, Italia}
\medskip \medskip

\begin{abstract}
Following recent literature on dS instability in presence of
interactions, we study the decay of massive particles in general FRW
models and the emission from naked singularities either
associated with $4D$ charged black holes or  $2D$ shock waves,   
by means of the Hamilton--Jacobi tunneling method. It is shown
that the two-dimensional semi-classical tunneling amplitude from a
naked singularity computed in that way is the same as the one-loop
result of quantum field theory.  
\end{abstract}


\section{Introduction}

It is well known  that  de Sitter (dS) space has gained tremendous
importance since the discovery by Riess and Perlmutter
\cite{Riess:1998cb,Perlmutter:1998np} that the Universe is -- against
any previous belief based on Einstein gravity with vanishing
cosmological constant -- in a current accelerating state.  

As far as this fundamental issue is concerned, two very interesting
papers on de Sitter  
space and the vacuum energy, one by A.~Polyakov \cite{Polyakov:2009nq}
and one by 
G.~Volovik \cite{Volo09} have  recently appeared. We recall that from
a classical point of view, it is 
generally believed that dS space is stable since: (i) it has a big
isometry group, namely $SO(1,4)$; (ii) linearized fluctuations of the
metric do not grow exponentially with time so that they are not able
to change the background dS metric; (iii) particles in dS space are
excitations over a dS invariant vacuum state (cfr. \cite{stability}).\\ 
Many authors have argued
\cite{Myhrvold:1983hx,Boyanovsky:2004ph,emil09,ugo,Volo09} that 
these observations are not sufficient to prove the classical stability
of dS space-time. In fact, whenever an interacting field theory is
present, dS space is unstable because of the non-vanishing probability
amplitude of massive particles radiating other massive particles. It
is worth stressing that the only force acting on the particles is due
to the gravitational background induced by $\Lambda$. Even if such
radiating process is small for small $\Lambda$ (as it is at present
time), no eternal dS space seems physically meaningful. Indeed, it is
expected that the particle production will eventually stop at the point
when back-reaction becomes significant. Even if it seems hard to make
precise predictions about what will occur at that point, Polyakov
\cite{Polyakov:2009nq} has
suggested that back-reaction will finally cancel any trace of the
Cosmological Constant and of dS space-time as well. The extension of
(some of) these results to general FRW models should give hope
that the deep analysis of Polyakov could be so extended. \\      
The decay of composite particles in dS space-time has been investigated
in \cite{Volo09} by means of semi-classical methods applied to a single
particle path. 
Very remarkably, the result obtained in this way turns out to be in
agreement with the asymptotic of an exact full QFT calculation given in
\cite{ugo}. The advantage of using the WKB approximation in favour of
exact QFT machinery is evident, since the extremely complicated
computations involved in \cite{ugo}.   

In this paper, first we would like to show that Volovik's result
\cite{Volo09} can be extended to a general FRW  
space-time, making use of the so-called Hamilton--Jacobi method
\cite{angh:2005,Kerner:2006vu,DiCriscienzo:2007fm}, implemented by  
using  the Kodama--Hayward invariant formalism \cite{Kodama,sean},
which can be applied to spherically symmetric space-times no matter if
static or dynamical. \\
Here the decay of the horizon corresponds to the existence
of a simple pole in the radial derivative of the action
while the particle's decay will correspond to a branch point
singularity in the radial particle's momentum.
Similar results will be obtained for static black holes possessing
time-like, or naked, singularities (a general reference for the study
of these objects is in Harada et al \cite{Harada:2001nj} and
references therein). They might even be produced in colliders in
certain brane world models \cite{Casadio:2001wh}.\\
Then we apply the null expansion method within the Hamilton--Jacobi
equation to see 
whether in a dynamical space-time region bounded by a naked singularity
there is radiation which can be interpreted as coming out from it. The
decay of the singularity, that if charged is eventually due to a 
screening by oppositely charged particles, or else its 
explosion \cite{Iguchi:2001ya}, is again
associated with a simple pole whose residue depends on which null
direction one is integrating the action's differential. As we will
see, this corresponds to the different components of the radiation's
quantum stress tensor. The fact that the Reissner-Nordstr\"{o}m
static singularity does not radiate (neutral particles) illustrates
how care has to be given when extrapolating two-dimensional results to
four-dimensional ones.     

The paper is organized in the following way: in \textsection 2 we
briefly resume the Kodama--Hayward approach to spherically symmetric
space-times; in \textsection 3 we apply the Hamilton--Jacobi method of
tunneling to FRW space-times; \textsection 4 is devoted to the
discussion of static spherically symmetric black-holes
endowed with time-like singularities; in \textsection 5 we analyze the
radiation from the singularity itself in a model of two-dimensional
dilaton gravity. Some conclusions will follow.

\section{The Kodama--Hayward formalism}
        
We recall that in  previous papers \cite{Hayward:2008jq,bob09} we
considered the quantum instability of dynamical black holes using a
variant of the tunneling method introduced by Parikh and Wilczek in
the static case to uncover aspects of 
back-reaction effects \cite{parikh}. These approaches are based on WKB
relativistic method (see, for example \cite{Visser} and  
more recently \cite{Menotti}) and only the leading terms of the
production rate probability are taken into account, leaving untouched
the pre-factor evaluation. Such evaluation, and possible back-reaction
effects, however, are not included in the present discussion which is
focused indeed, only on the leading WKB contribution to the production
rate.  
With  regard to the pre-factor issue, we limit ourselves  to mention Volovik's
arguments according to which the pre-factor is likely to vanish in the
case of horizon tunneling in dS space-time \cite{volo08}.\\ 
In our generalization of Volovik's calculation, the use of invariant
quantities plays a crucial role \cite{Hayward:2008jq,bob09}. In order
to illustrate them, let us recall that any spherically  
symmetric metric can locally be expressed in the form
\beq
\label{metric}
ds^2 =\gamma_{ij}(x^i)dx^idx^j+ R^2(x^i) d\Omega^2\,,\qquad i,j \in \{0,1\}\;,
\eeq
where the two-dimensional metric
\beq d\gamma^2=\gamma_{ij}(x^i)dx^idx^j
\label{nm}
\eeq
is referred to
as the normal metric, $\{x^i\}$ are associated coordinates and
$R(x^i)$ is the 
areal radius, considered as a scalar field in the two-dimensional
normal space. We recall that to have a truly dynamical solution,
i.e. to avoid Birkhoff's theorem, the space-time must be filled with
matter everywhere. Examples are the Vaidya solution, which contains a flux
of radiation at infinity, and FRW solutions which contain a perfect
fluid.\\  
A relevant scalar quantity in the reduced normal space is 
\beq
\chi(x)=\gamma^{ij}(x)\partial_i R(x)\partial_j R(x)\,, \label{sh} 
\eeq 
since the dynamical trapping horizon, if it exists, is located in
correspondence of  
\beq 
\chi(x)\Big\vert_H = 0\,,  \label{ho} 
\eeq
provided that $\partial_i\chi\vert_H \neq 0$.
The Misner--Sharp gravitational energy, in units $G=1$, is defined by
\beq
E_{MS}(x)=\frac{1}{2} R(x)\left[1-\chi(x) \right]\,. \label{MS}
\eeq
This is an invariant quantity on the normal space. Note also that, on
 the horizon, $E_{MS}\vert_H =\frac{1}{2} R_H$. Furthermore, one can
 introduce a dynamic surface 
gravity \cite{Hayward:2008jq} associated with this dynamical
horizon, given by the normal-space scalar 
\beq
\kappa_H=\frac{1}{2}\Box_{\gamma} R \Big\vert_H\,. \label{H} 
\eeq 
Recall that, in the spherical symmetric dynamical case, it is possible
to introduce 
the Kodama vector field $K$, with $(K^\alpha G_{\alpha\beta})^{;\beta}
=0$ that can be taken as its defining property. 
Given the metric (\ref{metric}), the Kodama vector components are
\beq 
K^i(x)=\frac{1}{ \sqrt{-\gamma}}\,\varepsilon^{ij}\partial_j R\,,
\qquad K^\theta=0=K^\varphi \label{ko} \;. 
\eeq 
The Kodama vector gives a preferred flow of time and in this sense it
generalizes the 
flow of time given by the Killing vector in the static case.
As a consequence, we may introduce the invariant energy associated
with a particle of mass $m$ by means of the scalar quantity on the
normal space 
\beq 
\label{e} 
\omega =- K^{i}\partial_i I\,, 
\eeq 
where $I$ is the particle action which we assume to satisfy the
reduced Hamilton--Jacobi equation  
\beq 
\label{hj} 
\gamma^{ij}\partial_i I \partial_j I + m^2=0\,. 
\eeq 
We shall call (\ref{e}) the Kodama, or generalized Killing energy.
As we allow for non-minimal gravitational coupling, the substitution
$m^2 \rightarrow m^2+\xi \mathcal R $ is in order whenever $\xi\neq
0$, $\mathcal R$ being the Ricci curvature scalar and $\xi$ a
dimensionless coupling constant. 

% We stress again the importance to have at disposal an invariant
% definition of energy. 

\section{ The FRW Space-time}

As a first application of the formalism, let us consider a generic FRW
space-time with constant curvature spatial sections. Its line element
can be written as 
\beq
ds^2=-dt^2+a^2(t)\frac{dr^2}{1-\hat k r^2}+ [a(t)r]^2 d \Omega^2 \;.
\label{frw nflat} 
\eeq
Here $\hat k :=\frac{k}{l^2}$, where $l$ is such that $a(t)l$ is
the curvature radius of 
the constant curvature  spatial sections at time $t$ and, as usual,
$k=0, -1, +1$ 
labels flat, open and closed three--geometries, respectively. In this
gauge, the normal reduced metric is diagonal and  
\beq
\chi(t,r)=1- [a(t)r]^2\left[H^2(t) + \frac{\hat k}{a^2(t)}\right]\,.
\eeq
The dynamical horizon is implicitly given by $\chi_H=0$, namely 
\beq 
R_H:= a(t) r_H = \frac{1}{\sqrt{ H^2(t)+\frac{\hat k}{a^2(t)}}}\,
,\qquad\mbox{with}\qquad  H(t)=\frac{\dot a(t)}{a(t)}\;, 
\label{h}
\eeq
provided the space-time energy density $\rho(t)$ is positive. The
surface $R_H(t)$ 
coincides with the Hubble radius as defined by astronomers for
vanishing curvature, but we shall call it 'Hubble radius' in any
case. The dynamical surface gravity is given by equation (\ref{H})
and reads 
\beq
\kappa_H= - \left( H^2(t) +\frac{1}{2} \dot H(t) + \frac{\hat
  k}{2a^2(t)}\right)\,R_H(t) <0 \,,\label{frw_dyn_sg}  
\eeq
and the minus sign refers to the fact the Hubble horizon is, in
Hayward's terminology, of the inner type. According to (\ref{ko}), the
Kodama vector is 
\beq\label{kv}
K= \sqrt{1-\hat k r^2}(\p_t -r H(t)\p_r )
\eeq 
so that the invariant Kodama energy of a particle is equal to  
\beq
\omega = \sqrt{1-\hat k r^2}(-\partial_t I + r H(t) \partial_r
I) \equiv \sqrt{1-\hat k r^2}\,\tilde{\omega}
 \label{b2}
\eeq 
Notice that $K$ is space-like for $ra>(H^2+\hat{k}/a^2)^{-1/2}$,
i.e. beyond the horizon. It follows that we can only ask for particles
to be emitted in the inner region, $r< r_H$.\\
The next ingredient is the reduced Hamilton--Jacobi equation for a relativistic
particle with mass parameter $m$,    
\beq
 -(\p_t I)^2 + \frac{(1-\hat k r^2) }{a^2(t)} \,(\p_r I)^2 + m^2=0\,.
\label{b}
\eeq
Making use of (\ref{b2}), one can solve for $\p_r I$, namely
\beq
\p_r I=-\frac{a H \tilde \omega (a r) \pm a\sqrt{\omega^2 -m^2 +
    m^2\,\left(H^2 + \frac{\hat k}{a^2}\right)\,(a r)^2}}{1-\left(H^2
  + \frac{\hat k}{a^2}\right) \,(a r)^2}\,,  
\label{g}
\eeq
with the signs chosen according to which direction we think the
particle is propagating.
The effective mass here defines two important and complementary energy
scales: if one is interested in the horizon tunneling then only
the pole matters (since the denominator vanishes), and we may neglect
to all the extents the mass parameter setting $m=0$ (since its
coefficient vanishes on the horizon). \\  
On the opposite, in investigating other effects in the bulk away from
the horizon, such as the decay rate of composite particles, the 
role of the effective mass becomes relevant as the energy of the particle
can be smaller than the energy scale settled by $m$, and the square
root can possibly acquire a branch cut singularity.     

\subsection{Horizon tunneling}

As an application of the last formula we may derive, following
\cite{bob09}, the cosmic 
horizon tunneling rate. To this aim, as we have anticipated, the
energy scale is such that near the  
horizon, we may neglect the particle's mass, and note that radially
moving massless particles follow a null direction. Then, along a null
radial direction from the horizon to the inner region, we have
\beq 
\delta t = -\frac{a(t)}{\sqrt{1-\hat{k}r^2}} \delta r. \label{ne}
\eeq
The outgoing particle action, that is the action for particles coming
out of the horizon towards the inner region, is then
\begin{eqnarray}
I &=& \int dt\,\p_t I + \int dr\,\p_r I  \label{minus}\\
&=&  2 \int dr \p_r I
\,\label{+action} 
\end{eqnarray}
upon solving the Hamilton--Jacobi equation \eqref{b} with zero mass and using
(\ref{ne}). For $\p_rI$ we use now Eq.~\eqref{g}, which exhibits a
pole at the vanishing of the function $F(r,t):=1-(a^2H^2+\hat k)r^2 $,
defining the horizon position. Expanding $F(r,t)$ again along a null
direction, one gets 
\beq
F(r,t) \approx  + 4 \kappa_H a(t) (r-r_H) +\dots \;,\label{hay sg}
\eeq
where $\kappa_H$ given in (\ref{frw_dyn_sg}) represents the
dynamical surface gravity associated with the horizon. \\ 
In order to deal with the simple pole in the
integrand, we implement Feynman's $i\epsilon$~{--}~prescription. In
the final result, beside a real (irrelevant) contribution, we 
obtain the following imaginary part \cite{bob09} 
\beq \Im\, I =-\frac{\pi \omega_H}{\kappa_H}\, .\label{im} 
\eeq
This imaginary part is usually interpreted as arising because of a
non-vanishing tunneling probability rate of (massless) particles
across the cosmological horizon, 
\beq
\Gamma \sim \exp\left(-2\Im \,I\right) \sim e^{-\frac{2 \pi}{(-
    \kappa_H)}\cdot \omega_H}. 
\eeq
Notice that, since $\kappa_H <0$ and $\omega_H >0$ for physical
particles, (\ref{im}) is positive definite. As showed in \cite{bob09},
this result is invariant since the quantities appearing in the
imaginary part are manifestly invariant. As a consequence, we may
interpret  $T=-\kappa_H/2 \pi$ as the dynamical temperature associated
with FRW space-times. In particular, this gives naturally a positive
temperature for de Sitter space-time, a long debated question years
ago, usually resolved by changing the sign of the horizon's energy.
It should be noted that in literature, the dynamical temperature is
usually given in the form  $T=\frac{H}{2\pi}$ (exceptions are the
papers \cite{Wu:2008ir}).  
Of course this is the expected result for dS space in inflationary
coordinates, but it ceases to be correct in any other coordinate
system. In this regard, the $\dot H$ and $\hat k$ terms are crucial in
order to get an invariant temperature. \\
The horizon's temperature and the ensuing heating of matter was 
foreseen several years ago in the interesting paper \cite{Brout:1987tq}.   

\subsection{Decay rate of unstable particles}

We are now ready to present the generalization of the result presented
in \cite{Volo09} for de Sitter space to a generic FRW 
space-time. Let us consider the decay rate
of composite particles in a regime where the energy of the decay product
is lower than their proper mass $m$. A crucial point is to
identify the energy of the particle before the decay with its Kodama
energy. After the decay process, we denote by $m$ the effective mass
parameter of one of the decay products (recall it may contain 
a curvature term). The relevant contribution to the action comes from
the radial momentum given by 
equation  (\ref{g}). If we introduce the instantaneous radius $r_0$ by
\beq
[a(t) r_0]^2 = R_0^2 := \left(1- \frac{\omega^2}{m^2}\right) R_H^2\;,
\label{ir} 
\eeq
where $R_H$ is the horizon radius given by Eq.~\eqref{h}, then
the classical forbidden region is  $0< r < r_0 $. Thus, from \eqref{g},
we see that for the unstable particle, say with mass $m_0$, sitting at
rest at the origin of the comoving coordinates, one has an imaginary
part of the action as soon as the decay product is tunneling into
this region to escape beyond $r_0$, 
\beq
\Im\, I= m R_H \int_0^{R_0} dR \frac{\sqrt{R_0^2 - R^2}}{R_H^2 - R^2}\,.
\eeq
The integral can be computed exactly at fixed $t$, and is
\beq
\Im\, I = \frac{\pi}{2} \,R_H \,(m-\omega) >0\,,
\eeq
leading to a rate which, assuming a two-particle decay, takes the form
\beq
\Gamma=\Gamma_0e^{-2\pi\,R_H\,(m-\omega)}   \,,
\label{g1}
\eeq 
where $\Gamma_0$ is an unknown pre-factor depending on the coupling
constant of the interaction which is causing the decay (for instance,
for a $\lambda\phi^3$ interaction one should have
$\Gamma_0\sim\lambda^2$.) Of course, each newly produced particle will
itself decay, leading possibly to the instability mechanism first
discussed by Myhrvold \cite{Myhrvold:1983hx} in dS space.\\  
Since the tunneling process locally conserves energy one
should put $\omega=m_0/2$, so that the tunnelled particle will emerge
in the classical region at $r=r_0$ with vanishing
momentum. Furthermore, the result is again invariant against
coordinate changes, since both $\omega$ and $R_H$ are invariantly
defined quantities. \\
A particularly interesting case is the de Sitter space-time. The line
element in the static patch reads 
\beq
ds^2=-(1- H_0^2r^2) dt^2+\frac{dr^2}{(1- H_0^2r^2 )}+ r^2 d \Omega^2\,, 
\label{ds s} 
\eeq
in the inflationary flat patch is
\beq
ds^2=-dt^2+ e^{2H_0t} d\,\overrightarrow x^2 \,,
\label{dsf} 
\eeq
while in global coordinates 
\beq
ds^2=-dt^2+ \cosh^2 (H_0t) d \Omega_3^2\,.
\label{dsg} 
\eeq
The so-called \textquotedblleft fluid'' static form discussed by
Volovik is instead 
\beq
ds^2=-dt^2+(dR-H_0 R dt)^2+ R^2 d \Omega^2\,.
\label{nolan nflat} 
\eeq
As already stated, a direct calculation leads always to $\kappa_H= -
H_0$ for the surface gravity and   
\beq
\Im\, I=\frac{\pi}{2H_0}(m-\omega)
\label{g11}
\eeq
for the imaginary part (\ref{g1}), independently by the coordinate
system in use. In the \textquotedblleft fluid'' gauge (\ref{nolan
  nflat}), putting 
$\omega=\frac{m_0}{2}$, the above result has been obtained by Volovik
\cite{Volo09}, in agreement with the exact result of \cite{ugo}.

\section{Black hole's singularities}

One may investigate if the method can be extended to the case
of static black holes.  
With regard to this, we consider the exterior region of a spherically
symmetric static 
black hole space-time and repeat the same argument. Quite generally, we
can write the line element as  
\beq
ds^2 = -  e^{2\psi(r)} C(r)dt^2 + C^{-1}(r) dr^2 + r^2
d\Omega^2.\label{clifton} 
\eeq
The radial momentum turns out to be,
\beq
\int dr\,\p_r I = \int dr\, \frac{\sqrt{\omega^2 - m^2 C(r)
    e^{2\psi(r)}}}{C(r)e^{\psi(r)}}\;.  
\eeq
The analysis of this integral is made easier by setting
$\omega=0$, which should correspond to particle creation: in fact,
according to the interpretation of the Kodama energy we 
gave before, this approximation simulates the vacuum condition. Then 
\beq
\int dr\,\p_r I =  m \int_{r_1}^{r_2} dr \frac{1}{\sqrt{-C(r)}}\,,
\label{inf} 
\eeq
where the integration is performed in every interval $(r_1,r_2)$ in
which $C(r) >0$.  
Equation (\ref{inf}) shows that, under very general conditions, in
static black hole space-times there could be a  decay rate whenever a
 region where $C(r)$ is positive exists.

As a first example,  let us analyze the Schwarzschild black hole. 
For the exterior (static) solution, one has  $C(r)=1-2M/r>0$ and
$\psi(r)=0$, thus the imaginary part diverges 
since the integral has an infinite range. We conclude that the
space-like singularity does not emit particles. In the interior, it is
possible to show that the Kodama vector is space-like, thus no energy
can be introduced. A similar conclusion has been obtained also for the
Big Bang cosmic singularity, the only scale factor leading to particle
emission being $a(t)\sim t^{-1}$. This is like a big rip in the past.
 
The situation is different when a naked singularity is
present. Consider a neutral particle in the Reissner--Nordstr\"{o}m
solution with mass $M$ and charge $Q>0$ (for definiteness) given by
the (spherically symmetric) line element  
\beq
ds^2=-\frac{(r-r_-)(r-r_+)}{r^2}dt^2+\frac{r^2}{(r-r_-)(r-r_+)}\,dr^2+r^2d
\Omega^2  \,.
\eeq
Here $r_{\pm}=M\pm\sqrt{M^2-Q^2}$ are the horizon radii connected to
the black hole mass and charge by the relations  
\beq
M=\frac{r_+ + r_-}{2}, \quad Q=\sqrt{r_+r_-}\,.
\eeq
The Kodama energy coincides with the usual Killing energy and
\beq
 C(r) = \frac{(r-r_-)(r-r_+)}{r^2}\;, \qquad \mbox{and}\qquad \psi(r)=0\;.
\eeq
% thus, equation (\ref{inf}) becomes,
% \[
% I=\int\sqrt{-m^2 C^{-1}(r)}dr
% \]
The metric function $C(r)$ is negative in between the two horizons,
so there the action is real. On the other hand it is positive within
the outer communication domain, $r>r_+$, but also within the region
contained by the inner Cauchy horizon, that is $0<r<r_-$. Thus,
because of (\ref{inf}) and assuming
the particles come created in pairs, we obtain
\beq
\label{ps}
\Im\, I=-m\int_0^{r_-}\frac{r}{\sqrt{(r_--r)(r_+-r)}}\;dr 
=  mQ-\frac{mM}{2}\ln\left(\frac{M+Q}{M-Q}\right)\;.
\eeq
Modulo the pre-factor over which we have nothing to say , there is a
probability 
\beq\label{p1}
\Gamma \sim \exp(-2\Im I) =
\left(\frac{M-Q}{M+Q}\right)^{mM}e^{-2Qm}\;. 
\eeq
Pleasantly, (\ref{p1}) vanishes in the extremal limit $M=Q$. Being computed
for particles with zero energy, we can interpret this as a particle
creation effect by the strong gravitational field near the
singularity. Since the electric field is of order $Q/r^2$ near $r=0$,
there should also be a strong Schwinger's effect. In that case one
should write the Hamilton--Jacobi equation for charged particles.  It
would be an 
interesting question whether this effect could lead to a screening
of the singularity, and ultimately to its disappearance. 

Next we consider the hairy black hole solution in the Jordan frame, for
$\Lambda=\frac{3}{\ell^2} > 0$. This is a black hole solution in the
Einstein theory with a self-interacting and  
conformally coupled scalar field (see\cite{t,mario}). One has 
\beq
C(r)= -\frac{r^2}{\ell^2}+\frac{(r+r_0)^2}{r^2}\,, \qquad
\mbox{and}\qquad  \psi(r)=0\,,  
\eeq
while that scalar field is
\beq
\phi(r)=\sqrt{6}\,\frac{r_0}{r+r_0}\,.
\eeq
For our purposes, $r_0=-M$ is the interesting case, since then one
has four real roots of $C(r)=0$, the first positive is 
the inner horizon
\beq
r_+=\frac{\ell}{2}\left( \sqrt{1+\varepsilon}-1\right)\,,  
\eeq
the second root represents the event horizon
\beq
r_H=\frac{\ell}{2}\left(1- \sqrt{1-\varepsilon}\right)\,,  
\eeq
the third root is associated with the cosmological horizon
\beq
r_C=\frac{\ell}{2}\left(1+ \sqrt{1-\varepsilon}\right)\,,  
\eeq
and the negative root is
\beq
r_{-}=-\frac{\ell}{2}\left( \sqrt{1+\varepsilon}+1\right)\,,  
\eeq
where $\varepsilon=\frac{4M}{\ell}$. The quantity $C(r)$ is positive
in the two static regions $0 < r< r_+$ and  $r_H < r< r_C$. 
Thus, there is a naked singularity in the first region at $r=0$ where
-- with specific choice of the sign, 
\beq\label{ps1}
\Im\, I=m
\ell\int_0^{r_+}\frac{r}{\sqrt{(r_+ - r)(r_H - r)(r_C - r)(r - r_{-})}}\,dr\,. 
 \eeq
% Re-scaling $r\rightarrow x:= r/l$, the same integral becomes,
% \beq
% \mbox{Im}\, I=m
% \int_0^{x_+}\frac{x}{\sqrt{(x_+-x)(x-x_H)(x-x_C)(x-x_{-})}}dr\,. % \eeq
The integral is the sum of two special functions, the confluent
hypergeometric function and an elliptic integral of the  
third kind. However if the quantity $\varepsilon=\frac{4M}{\ell}$ is
small, a direct calculation leads to 
\beq
\label{ps2}
\Im\, I \simeq  m \ell
\int_0^{b-a}\frac{r}{(r-b)\sqrt{\ell^2-(r+b)^2}}\,dr\,,
\eeq
where $b=\frac{\ell \varepsilon}{4}=M$ and
$a=\frac{\ell\varepsilon^2}{16}=\frac{M^2}{\ell}$. This integral can
be evaluated, and the corresponding  leading terms are 
\beq
\Im\, I \simeq   m M\left[1 + \ln
  \left(\frac{M}{\ell}\right)\right]+ O(\varepsilon^2)\,. 
\eeq
Again, modulo the pre-factor and assuming the particles are created in
pairs, the leading term of production rate is 
\beq\label{p2}
\Gamma\sim \exp(-2\Im I)=\left(\frac{M}{\ell}\right)^{-2 m M}
e^{-2 m M}\,.
\eeq
It is worth to mention that, given $\frac{M}{\ell}\ll 1$,  still the
order of magnitude of (\ref{p2}) strongly depends on the reciprocal
relation between the parameters $m, M, \ell$.\\  
All these results can be interpreted as particle creation effects by
the bulk gravitational field. In fact the probabilities given by
Eq.~s\eqref{p1}, \eqref{p2}, do not depend on the position of the
creation event, so actually the full amplitude associated to a
space-time region must be proportional to the four-volume of the
region. Equivalently, the formulas give creation probability per unit
volume per unit time. \\
A complementary and potentially interesting effect is the emission from
the naked singularity itself. We investigate this problem in the
following section for the case of $2D$ dilaton gravity, and return to
RN afterwards.    

\section{Radiation from a naked singularity}

Consider the following metric \cite{vaz1993}
\beq
\label{ns}
ds^2=\sigma^{-1}dx^+dx^-, \qquad
\sigma=\lambda^2x^+x^--a(x^+-x_0^+)\theta(x^+-x_0^+) 
\eeq
where $\lambda$ is related to the cosmological constant by
$\Lambda=-4\lambda^2$. 
This metric arises as a solution of $2D$ dilaton gravity coupled to a
bosonic field with stress tensor $T_{++}=2a\delta(x^+-x_0^+)$, 
describing a shock wave. A look at Fig.~1
reveals that $\sigma=0$ is a naked singularity partly to the future of
a flat space region, usually named the linear dilaton vacuum. The
heavy arrow represents the history of the shock wave responsible for
the existence of the 
time-like singularity. The
Hamilton--Jacobi equation implies either $\partial_+I=0$ or 
$\partial_-I=0$, $I$ being the action. To find the ingoing flux we
integrate along $x^+$ till we encounter the naked
singularity, using $\partial_-I=0$, so that
\beq
I=\int dx^+\partial_+I = \int\omega \frac{dx^+}{2\sigma}
=\int\frac{\omega dx^+}{2(\lambda^2x^--a)(x^++ax_0^+/C-i\epsilon)}   
\eeq
where $C = C(x^-):=(\lambda^2x^- - a)$ and
$\omega=2\sigma\partial_+I$ is the familiar Kodama's energy; note the Feynman
$i\epsilon$--prescription. Thus the imaginary part immediately
follows\footnote{Using 
  $(x-i\epsilon)^{-1}=P\frac{1}{x}+i\pi\delta(x)$.}, giving the
absorption probability as a function of retarded time 
\beq
\Gamma(\omega)=\Gamma_0e^{-2\Im I}=\Gamma_0e^{-\pi\omega/C(x^-)}
\eeq
$\Gamma_0$ being some pre-factor of order one.

\begin{figure}[h]
\begin{center}\includegraphics[scale=0.7]{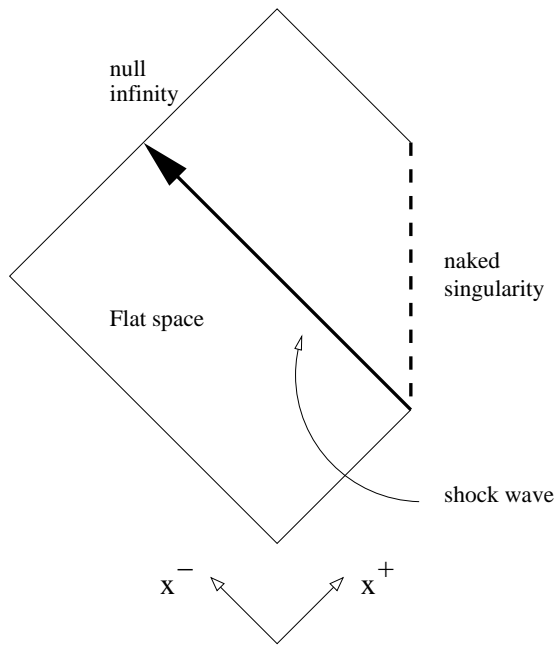}\end{center}
\caption{The naked singularity formed by the shock wave.}
\end{figure}

The flux is computed by
integrating the probability over the coordinate frequency
$\tilde{\omega}=\omega/\sigma$, with the density of states measure
$\frac{1}{2\pi}d\tilde{\omega}$, giving 
\beq\label{iflx}
T_{++}=\frac{\Gamma_0}{2\pi}\int\Gamma(\sigma\tilde{\omega})\tilde{\omega}
d\tilde{\omega}=\Gamma_0\frac{(\lambda^2x^--a)^2}{2\pi^3\sigma^2}\;.
\eeq 
Similarly, in order to find the outgoing flux we
integrate along $x^-$ starting from the naked singularity, this time
using $\partial_+I=0$. A similar calculation first gives 
\beq
\Im I=\frac{\pi\omega}{2\lambda^2x^+}\;.
\eeq
Then, integrating the probability over the coordinate frequency, the
outgoing flux 
\beq\label{oflx}
T_{--}=\Gamma_0\frac{\lambda^4 (x^+)^2}{2\pi^3\sigma^2}
\eeq
is obtained (strictly speaking the outgoing flux would be
$2T_{++}-2T_{--}$). The conservation equations  
\begin{eqnarray}
\sigma\partial_+T_{--}+\partial_-(\sigma T_{+-}) &=& 0 \nonumber \;,\\
\sigma\partial_-T_{++}+\partial_+(\sigma T_{+-})&=& 0\,,
\end{eqnarray}
will determine the components only up to arbitrary functions $B(x^-)$
and $A(x^+)$, respectively, corresponding to the freedom of the choice
of a vacuum. For instance, requiring the fluxes to vanish in the linear
dilaton vacuum fixes them uniquely. As regards $T_{+-}$, it is well known that 
it is given by the conformal anomaly: $T=4\sigma T_{+-}=R/24\pi$ (for
one bosonic d.o.f.). Matching to the anomaly gives the pre-factor
$\Gamma_0=\pi^2/24$, of order one indeed. These results agree with the
one-loop calculation to be found in \cite{vaz1993}. Note that the
stress tensor diverges approaching the singularity, indicating that
its resolution will not be possible within classical gravity but
requires instead quantum gravity \cite{Harada:2000me,Iguchi:2001ya}. \\ 

We return now to the RN solution. Could it be that the naked
singularity emitted particles? In the $4D$ case one easily sees that
the action has no imaginary part along null trajectories either ending
or beginning at the singularity. Formally this is because the Kodama
energy coincides with the Killing energy in such a static manifold and
there is no infinite redshift from the singularity to infinity. Even
considering the metric as a genuinely two-dimensional solution, this
would lead to an integral for $I$ 
\beq
I=\int\frac{\omega(r-r_+)(r-r_-)}{r^2}\,dx^+
\eeq
where $x^{\pm}=t\pm r_*$, with
\beq
r_*=r+\frac{r_+^2}{r_+-r_-}\ln[(r_+-r)/r_+]-\frac{r_+^2}{r_+-r_-}
\ln[(r_--r)/r_-]=\frac{x^+-x^-}{2}    \;.
\eeq
But close to the singularity
\beq
r^2=(3r_+r_-/2)^{2/3}(x^+-x^-)^{2/3}+\cdots
\eeq
not leading to a simple pole. It is fair to say that the RN naked
singularity will not emit particle in this approximation.\\
This seems to be coherent with QFT results. Fig. 2 depicts part of a Penrose's
diagram for the region near the singularity of RN (the left one, say).
\begin{figure}[h!]
\begin{center}
\includegraphics[scale=0.7]{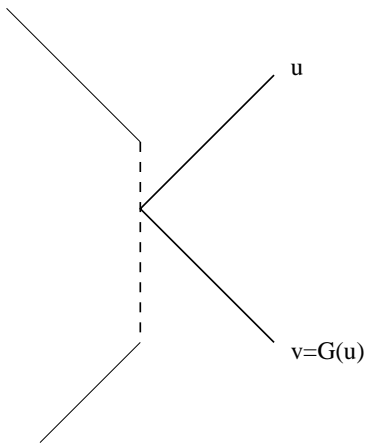}
\caption{Part of the Penrose diagram of the RN region close to the
  (left) time-like singularity.} 
\end{center}
\end{figure}
With the customary $u=x^-$ and $v=x^+$, the map $u\to v=G(u)$ gives
that ingoing null geodesics which after reflection in the origin
emerges as the outgoing geodesic drawn. According
to \cite{Ford:1978ip}, the radiated $s$--wave power of a minimally coupled
scalar field is given in terms of the map $G(u)$ and its derivatives,
by the Schwarzian derivative 
\beq
W=\frac{1}{24\pi}\left[\frac{3}{2}\left(\frac{G^{''}}{G^{'}}\right)^2-
\frac{G^{'''}}{G^{'}}\right]\;.
\eeq
The $(u,v)$ section of the RN metric is conformally flat, hence the
above map is trivial (or linear) and $W=0$.

\section{Conclusions}

In this paper, several applications of the tunneling method have been presented. 
In our opinion, the most pleasant aspect of the semi-classical tunneling
method in the analyzed context is its flexibility and the wide range of
situations to which it can be applied. Normally great efforts are
needed to analyze quantum effects in gravity, while the tunneling
picture promptly gives strong indications of what could happen. The
obtained agreement between the particle decay rates from tunneling
methods with the asymptotic of the exact results, when they exist in
particular backgrounds like dS space, gives confidence of their
validity in more general situations. Similarly, the coincidence of the
tunneling radiation from naked singularities with one-loop quantum
field theory results gives confidence that similar effects also exists
for naked singularities in $4D$ backgrounds. In particular we have
shown that the same expression derived from the Hamilton--Jacobi
equation can handle several quantum effects: radiation from dynamical
horizons, both cosmological and collapsing, gravitational
enhancement of particle decay which would otherwise be forbidden by
conservation laws, and radiation from naked singularities, at least in
some $2D$ dilaton gravity models.

\ack{The authors thank G.~Volovik, R.~Casadio, G.~Venturi for useful
  discussions.}


\begin{thebibliography}{99}



\bibitem{Riess:1998cb}
  A.~G.~Riess {\it et al.} 
  Astron.\ J.\  {\bf 116}, 1009 (1998).

\bibitem{Perlmutter:1998np}
  S.~Perlmutter {\it et al.} 
  Astrophys.\ J.\  {\bf 517}, 565 (1999).

\bibitem{Polyakov:2009nq}
  A.~M.~Polyakov,
  %``Decay of Vacuum Energy,''
arXiv:0912.5503 [hep-th].

\bibitem{Volo09}
   G.~E.~Volovik, JETP Lett.\  {\bf 90}, 1 (2009).

\bibitem{stability}
 G.~Borner and H.~P.~Durr, Il Nuovo Cimento, Vol. LXIV A, No. 3
 (1969); R.~Bousso, A.~Maloney and A.~Strominger, Phys. Rev. {\bf
 D65}, 104039 (2002); Y.~B.~Kim, C.~Y.~Oh and N.~Park,
 arXiv:hep-th/0212326; E.~Lifshitz, J. Phys. (USSR) {\bf 10},116
 (1946); E.~Mottola, Phys. Rev. {\bf D31}, 754 (1985).  

\bibitem{Myhrvold:1983hx}
  N.~P.~Myhrvold,
  %``Runaway Particle Production In De Sitter Space,''
  Phys.\ Rev.\  D {\bf 28}, 2439 (1983).

\bibitem{Boyanovsky:2004ph}
  D.~Boyanovsky, H.~J.~de Vega and N.~G.~Sanchez,
  %``Particle decay during inflation: Self-decay of inflaton quantum
  %fluctuations during slow roll,''
  Phys.\ Rev.\  D {\bf 71}, 023509 (2005).
%   [arXiv:astro-ph/0409406].

\bibitem{emil09}
  A.~M.~Polyakov, Nucl.\ Phys.\ {\bf B797} (2008), 199;
  E.~T.~Akhmedov, P.~V.~Buividovich and D.~A.~Singleton, arXiv:
  0905.2742 [gr-qc];   
  E.~T.~Akhmedov, arXiv: 0909.3722 [hep-th].

\bibitem{ugo}
  J.~Bros, H.~Epstein and U.~Moschella,
  %``Lifetime of a massive particle in a de Sitter universe,''
  JCAP {\bf 0802}, 003 (2008) ;
%   [arXiv:hep-th/0612184];
  J.~Bros, H.~Epstein and U.~Moschella,
%   ``Particle decays and stability on the de Sitter universe,''
  arXiv: 0812.3513 [hep-th];
  J.~Bros, H.~Epstein, M.~Gaudin, U.~Moschella and V.~Pasquier,
%   ``Triangular invariants, three-point functions and particle stability on the de Sitter universe,''
  arXiv: 0901.4223 [hep-th].

\bibitem{angh:2005}
%HAWKING RADIATION AS TUNNELING FOR EXTREMAL AND ROTATING BLACK HOLES.
M. Angheben, M. Nadalini, L. Vanzo and S. Zerbini,
JHEP {\bf 0505}, 014 (2005);
% HAWKING RADIATION AS TUNNELING: THE D DIMENSIONAL ROTATING CASE
M. Nadalini, L. Vanzo and S. Zerbini,
  J. Physics A: Math. Gen. {\bf 39}, 6601 (2006).

\bibitem{Kerner:2006vu}
  R.~Kerner and R.~B.~Mann,
  %``Tunnelling, Temperature and Taub-NUT Black Holes,''
  Phys.\ Rev.\  D {\bf 73}, 104010 (2006).
%   [arXiv:gr-qc/0603019].


\bibitem{DiCriscienzo:2007fm}
 R.~Di Criscienzo, M.~Nadalini, L.~Vanzo, S.~Zerbini and G.~Zoccatelli,
  %``On the Hawking radiation as tunnellingfor a class of dynamical black
  %holes,''
  Phys.\ Lett.\ {\bf B657}, 107 (2007).

\bibitem{Kodama}
H.~Kodama,
  %``Conserved Energy Flux For The Spherically Symmetric System And The Back
  %Reaction Problem In The Black Hole Evaporation,''
  Prog.\ Theor.\ Phys.\  {\bf 63}, 1217 (1980).

\bibitem{sean}
S.~A.~Hayward,
  %``Unified first law of black-hole dynamics and relativistic
 %thermodynamics,''
  Class.\ Quant.\ Grav.\  {\bf 15}, 3147 (1998).

\bibitem{Hayward:2008jq}
  S.~A.~Hayward, R.~Di Criscienzo, L.~Vanzo, M.~Nadalini and S.~Zerbini,
  %``Local Hawking temperature for dynamical black holes,''
  Class.\ Quant.\ Grav.\  {\bf 26 }, 062001 (2009).

\bibitem{bob09}
  R.~Di Criscienzo, S.~A.~Hayward, M.~Nadalini, L.~Vanzo and S.~Zerbini,
   Class.\ Quant.\ Grav.\  {\bf 27}, 015006 (2010).
%   arXiv:0906.1725 [gr-qc].

\bibitem{Harada:2001nj}
  T.~Harada, H.~Iguchi and K.~i.~Nakao,
  %``Physical processes in naked singularity formation,''
  Prog.\ Theor.\ Phys.\  {\bf 107}, 449 (2002).
%   [arXiv:gr-qc/0204008].

\bibitem{Casadio:2001wh}
  R.~Casadio and B.~Harms,
  %``Can black holes and naked singularities be detected in accelerators?,''
  Int.\ J.\ Mod.\ Phys.\  A {\bf 17}, 4635 (2002)
  [arXiv:hep-th/0110255].


\bibitem{Iguchi:2001ya}
  H.~Iguchi and T.~Harada,
  %``Physical aspects of naked singularity explosion - How does a naked
  %singularity explode?,''
  Class.\ Quant.\ Grav.\  {\bf 18}, 3681 (2001)
  [arXiv:gr-qc/0107099].


\bibitem{parikh}
M.~K.~Parikh and F.~Wilczek,
%``Hawking Radiation as Tunneling,''
Phys.\ Rev.\ Lett.\  {\bf 85}, 5042 (2000).

\bibitem{Visser}
M.~Visser,
  %``Essential and inessential features of Hawking radiation,''
  Int.\ J.\ Mod.\ Phys.\ {\bf D12}, 649 (2003);
A.~B.~Nielsen and M.~Visser,
  %``Production and decay of evolving horizons,''
  Class.\ Quant.\ Grav.\ {\bf 23}, 4637 (2006).

\bibitem{Menotti}
  P.~Menotti,
  %``On the semiclassical treatment of Hawking radiation,''
  arXiv: 0911.4358 [hep-th].

\bibitem{volo08}
  G.~E.~Volovik,
%   ``On de Sitter radiation via quantum tunnelling,''
  arXiv: 0803.3367 [gr-qc].

\bibitem{Wu:2008ir}
  S.~F.~Wu, B.~Wang, G.~H.~Yang and P.~M.~Zhang,
  %``The generalized second law of thermodynamics in generalized gravity
  %theories,''
  Class.\ Quant.\ Grav.\  {\bf 25}, 235018 (2008).
%   [arXiv:0801.2688 [hep-th]].


\bibitem{Brout:1987tq}
  R.~Brout, G.~Horwitz and D.~Weil,
  %``ON THE ONSET OF TIME AND TEMPERATURE IN COSMOLOGY,''
  Phys.\ Lett.\  B {\bf 192}, 318 (1987).
  %%CITATION = PHLTA,B192,318;%%


\bibitem{t}
  C.~Martinez, R.~Troncoso and J.~Zanelli,
  %``De Sitter black hole with a conformally coupled scalar field in  four
  %dimensions,''
  Phys.\ Rev.\ {\bf  D67}, 024008 (2003).
%   arXiv:hep-th/0205319].

\bibitem{mario}
  M.~Nadalini, L.~Vanzo and S.~Zerbini,
  %``Thermodynamical properties of hairy black holes in n spacetimes
  %dimensions,''
  Phys.\ Rev.\ {\bf  D77}, 024047 (2008).
%   [arXiv:0710.2474 [hep-th]].

\bibitem{vaz1993}
C.~Vaz and L.~Witten, Phys.Lett. {\bf B325}, 27 (1994).
%[arXiv: hep-th/9311133] 

\bibitem{Harada:2000me}
  T.~Harada, H.~Iguchi, K.~i.~Nakao, T.~P.~Singh, T.~Tanaka and C.~Vaz,
  %``Naked singularities and quantum gravity: Interpreting the quantum
  %divergence in spherical collapse,''
  Phys.\ Rev.\  D {\bf 64}, 041501 (2001).
%   [arXiv:gr-qc/0010101].


\bibitem{Ford:1978ip}
  L.~H.~Ford and L.~Parker,
  %``Creation Of Particles By Singularities In Asymptotically Flat
  %Space-Times,''
  Phys.\ Rev.\  D {\bf 17}, 1485 (1978).
  %%CITATION = PHRVA,D17,1485;%%
\end{thebibliography}
\end{document}